\documentclass[Proceedings,11pt,letterpaper]{ascelike}
\usepackage[pdftex]{graphicx}
\usepackage{times}
\usepackage{amsfonts,amssymb,amsmath,amstext}
\begin{document}
\DeclareGraphicsExtensions{.jpg,.pdf,.eps}

\setlength{\textwidth}{6in}
\setlength{\oddsidemargin}{.25in}
\thispagestyle{empty}

\begin{figure}
 \flushright
 \scalebox{0.65}{\includegraphics{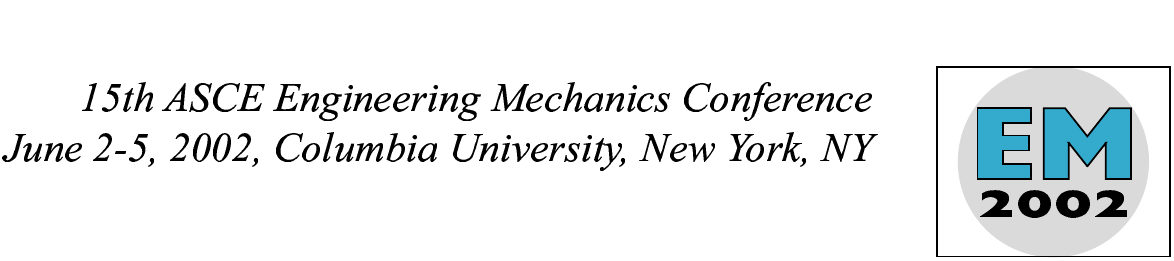}}
 \vspace*{-1in}
\end{figure}

\title{MICROMECHANICS-BASED PREDICTION OF THERMOELASTIC PROPERTIES OF HIGH ENERGY MATERIALS}
\author{
Biswajit Banerjee %
\thanks
{Dept. of Mechanical Engrg., University of Utah, Salt Lake City, UT 84112.
 E-mail : banerjee@eng.utah.edu.}
and
Daniel O. Adams %
\thanks
{Dept. of Mechanical Engrg., University of Utah, Salt Lake City, UT 84112.
 E-mail : adams@eng.utah.edu.}
}
\maketitle
\begin{abstract}
  High energy materials such as polymer bonded explosives are commonly used as
  propellants.  These particulate composites contain explosive crystals suspended
  in a rubbery binder.  However, the explosive nature of these materials
  limits the determination of their mechanical properties by experimental means.
  Therefore micromechanics-based methods for the determination of the effective 
  thermoelastic properties of polymer bonded explosives are investigated in 
  this research.
  Polymer bonded explosives are two-component particulate composites
  with high volume fractions of particles (volume fraction $>$ 90\%) and high
  modulus contrast (ratio of Young's modulus of particles to binder of 5,000-10,000).
  Experimentally determined elastic moduli of one such material, PBX 9501, are
  used to validate the micromechanics methods examined in this research.  The
  literature on micromechanics is reviewed; rigorous bounds on effective elastic
  properties and analytical methods for determining effective properties are
  investigated in the context of PBX 9501.
  Since detailed numerical simulations of PBXs are computationally expensive,
  simple numerical homogenization techniques have been sought.  Two such techniques
  explored in this research are the Generalized Method of Cells and the Recursive
  Cell Method.  Effective properties calculated using these methods have been 
  compared with finite element analyses and experimental data.  
\end{abstract}
%
%
\section{Introduction}
  Large scale simulations of the mechanical response of containers filled with 
  high energy (HE) materials require knowledge of the mechanical properties of 
  the materials involved.  Though mechanical properties of HE materials can be 
  determined experimentally, hazards associated with experiments on these 
  materials, as well as the attending costs, make this option unattractive.  As 
  computational capabilities have grown and improved numerical techniques developed, 
  numerical determination of the properties of HE materials has become possible.  In 
  this research, we explore micromechanics-based numerical methods for the 
  determination of the mechanical properties of HE materials, specifically polymer 
  bonded explosives (PBXs).  The polymer bonded explosive PBX 9501 has been 
  characterized experimentally and thus provides a basis for validating numerical 
  calculations.  

  PBXs provide unique challenges for micromechanical modeling.  These materials
  are viscoelastic particulate composites, contain high volume fractions of  
  particles, and the modulus contrast between the particles and the binder is
  extremely high.  For example, PBX 9501 contains about 92\% by volume of 
  particles and the modulus contrast between particles and binder of 20,000 at 
  room temperature and low strain rates.  In this research it is assumed that PBXs 
  are two-component particulate composites with the particles perfectly bonded to 
  the binder.  The components of PBXs are assumed to be isotropic and linear elastic, 
  and only the prediction of elastic moduli and coefficients of thermal expansion (CTEs) of 
  PBXs is addressed.  Finite element (FEM) analyses are used to compute the effective
  properties of PBX 9501 models and compared with properties computed using the 
  Generalized Method of Cells (GMC).  A new technique called the Recursive Cell Method
  (RCM) has been developed to address the limitations of GMC.  Effective properties
  of PBX 9501 are also calculated using RCM for some microstructures simulating PBX 9501.
\section{Polymer Bonded Explosives}
  Polymer bonded explosives are mixtures containing high volume fractions of stiff
  explosive crystals suspended in a continuous compliant binder.  Some common 
  PBXs~\cite{Gibbs80} are shown in Table~\ref{tab1}.
  \begin{table}[htb]
     \caption{Common polymer bonded explosives and their components.}
     \label{tab1}
     \centering
     \small
     \renewcommand{\arraystretch}{1.25}
     \begin{tabular}{l||c|c|c||c|c|c}
	\hline\hline
	PBX & Crystal & Weight & Volume & Binder & Weight & Volume \\
	    &         & (\%)   & (\%)   &        & (\%) & (\%) \\
        \hline
	PBX 9501 & HMX & 95 & 92 & Estane 5703 + BDNPA/F & 5 & 8 \\
	PBX 9010 & RDX & 90 & 87 & KEL-F-3700 & 10 & 13 \\
	PBX 9502 & TATB & 95 & 90 & KEL-F-800 & 5 & 10 \\
	\hline\hline
     \end{tabular}
     \normalsize
  \end{table}

  \subsection{PBX 9501}
  PBX 9501 is a mixture of HMX particles coated by a binder composed of Estane and a 
  plasticizer (BDNPA/F).  The HMX crystals are in the stable $\beta$ phase, have a monoclinic 
  structure~\cite{Skid97}, and are linearly elastic at room temperature.  The binder in 
  PBX 9501 is viscoelastic.  However at room temperature and low strain rates, the response 
  of the binder is close to linear elastic.  The composite (PBX 9501) is also linear elastic 
  at room temperature and low strain rates.  The elastic properties of PBX 9501 and it's 
  components at 23$^o$C and a strain rate of 0.05/s~\cite{Gray98,Zaug98} are shown 
  in Table~\ref{tab2}.
  \begin{table}[htb]
     \caption{Elastic properties of PBX 9501, HMX and binder.}
     \label{tab2}
     \centering
     \small
     \renewcommand{\arraystretch}{1.25}
     \begin{tabular}{l|c|c|c}
	\hline\hline
	Material & Young's Modulus  & Poisson's Ratio & CTE \\
		 & (GPa)            &           & ($\times 10^{-5}$/K) \\
        \hline
	PBX 9501 & 1.03  & 0.35 & 12 \\ 
	HMX      & 15.3  & 0.32 & 11.6 \\ 
	Binder   & 0.001 & 0.49 & 20   \\
	\hline\hline
     \end{tabular}
     \normalsize
  \end{table}
\section{Micromechanics Methods}
  The term ``micromechanics'' describes a class of methods that use continuum mechanics
  for determining the effective material properties of composites given the material properties 
  of the constituents.  Methods of determining the effective elastic properties and 
  coefficients of thermal expansion of particulate composites, containing up to 70\% by volume 
  of particles, are well established~\cite{Hashin83,Markov00,Burya01}.  Rigorous bounds on 
  effective properties~\cite{Milton02} and various analytical methods for calculating 
  effective properties~\cite{Torquato02} are also available.  However, these techniques 
  are not particularly useful for high volume fraction ($>$ 90\%) and high modulus contrast 
  composites such as PBX 9501.  Therefore numerical techniques are essential to the solution 
  of micromechanics problems for polymer bonded explosives.

  \subsection{Rigorous Bounds and Analytical Methods}
  Hashin-Shtrikman~\cite{Hashin63} and third order~\cite{Milton81} bounds have been
  calculated for the effective elastic moduli of PBX 9501.  The Rosen-Hashin 
  bounds~\cite{Rosen70} on the effective coefficient of thermal expansion of PBX 9501
  have also been calculated.  These bounds are shown in Table~\ref{tab3}.  The bounds on 
  the elastic moduli are quite far apart and hence of no practical use.  However, the 
  bounds on the CTE are within 1\% of each other.
  \begin{table}[htb]
     \caption{Bounds on the effective properties of PBX 9501.}
     \label{tab3}
     \centering
     \small
     \renewcommand{\arraystretch}{1.25}
     \begin{tabular}{l|c|c|c|c|c}
	\hline\hline
	& Expt. &\multicolumn{2}{c|}{Hashin-Shtrikman Bounds}&
		\multicolumn{2}{c}{Third Order Bounds}\\\cline{3-6}
	&       & Upper & Lower & Upper & Lower \\
	\hline
        Bulk Modulus (GPa) & 1.1 & 11.4 & 0.15 & 11.3 & 0.22 \\
        Shear Modulus (GPa) & 0.4 & 5.3  & 0.01 &  5.0 & 0.07 \\
	CTE ($\times 10^{-5}$/K) & 12 & 12.3 & 11.6 & - & - \\
	\hline\hline
     \end{tabular}
  \end{table}

  Analytical methods of interest for high volume fraction particulate composites include the 
  composite spheres assemblage~\cite{Hashin62}, the self-consistent scheme (SCS) ~\cite{Berry96}, 
  and the differential effective medium (DEM) approach~\cite{Markov00}.  Each of these 
  methods makes simplifying assumptions about the microstructure of the composites.  The 
  CTE can be calculated given knowledge of the isotropic bulk modulus of the 
  composite~\cite{Rosen70}.  The effective elastic moduli of PBX 9501 calculated using SCS
  and DEM, and the corresponding CTEs are shown in Table~\ref{tab4}.  It can be 
  observed that the analytical methods do not predict effective elastic moduli that are
  close to the experimental values.  However, the predicted effective CTE is close enough 
  to the experimental value to be of use and no further numerical calculation is required for 
  this property.
  \begin{table}[htb]
     \caption{Effective properties of PBX 9501 from analytical models.}
     \label{tab4}
     \centering
     \small
     \renewcommand{\arraystretch}{1.25}
     \begin{tabular}{l|c|c|c}
	\hline\hline
        & Bulk Modulus & Shear Modulus & CTE \\
	                              & (GPa)  & (GPa)  & ($\times 10^{-5}$/K) \\
	\hline
	Experiment                    & 1.1   & 0.4   & 12 \\
	\hline
	Self-Consistent Scheme        & 11.0  & 4.7   & 12.9 \\
	Differential Effective Medium & 0.2   & 0.08  & 12.5 \\
	\hline\hline
     \end{tabular}
  \end{table}
     
  \subsection{Numerical Approximations}
  The effective elastic moduli of a composite can be determined approximately by solving 
  the governing differential equations using numerical methods.  This process involves the 
  determination of a Representative Volume Element (RVE), the choice of appropriate boundary 
  conditions and the solution of the resulting boundary value problem. The effective stiffness 
  tensor ($C^*_{ijkl}$) of the composite is then calculated from the relation
  \begin{equation}
	\int_V \sigma_{ij}~dV = C^*_{ijkl} \int_V \epsilon_{kl}~dV
  \end{equation}
  where $V$ is the volume of the RVE, $\sigma_{ij}$ are the stresses, and $\epsilon_{kl}$
  are the strains.

  Finite element analysis is the most commonly used numerical technique use to determine
  effective properties.  The difficulties involved in discretization and solution of actual
  particulate geometries has led to simulations of simple geometries such as square or 
  hexagonal particles, mostly in two dimensions.  With improvement in computational
  capabilities random distributions of particles are being simulated more frequently~\cite{Mish01}.  
  The large computational cost associated with finite element 
  analyses and the poor accuracy in areas of high stress gradient have led to the exploration 
  of alternative methods of calculating effective properties.  Discrete spring network 
  models~\cite{Day92}, integral equation based methods~\cite{Greeng98} and Fourier transform
  based methods~\cite{Michel01} show the most promise.  However, a large amount of computation 
  is still required for convergence in all these methods.

  \subsubsection{Generalized Method of Cells (GMC)}
\begin{figure}[htb]
     \centering
     \scalebox{0.40}{\includegraphics{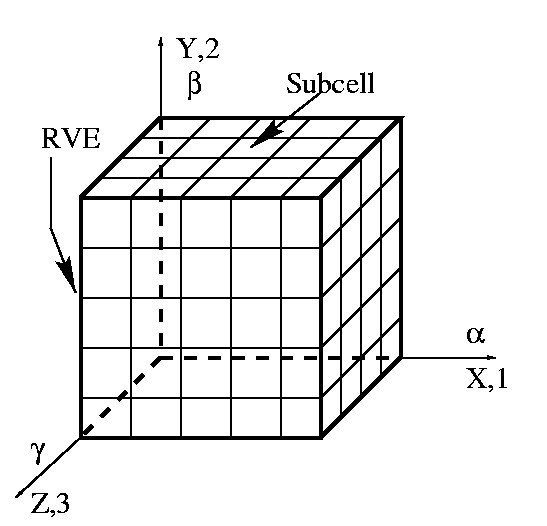}}
     \caption{Schematic of the Generalized Method of Cells.}
     \label{fig:GMC}
  \end{figure}

  The Generalized Method of Cells (GMC)~\cite{Aboudi96} has been used to model the 
  micromechanical behavior of various types of composites with relative success.  The advantage 
  of this method over other numerical techniques is that the full set of effective elastic 
  properties can be calculated in one step.  In this technique, the RVE is discretized 
  into a number of subcells as shown in Figure~\ref{fig:GMC}.  Continuity of displacements
  between subcells and subcell equilibrium are satisfied in an average sense using 
  integrals over subcell boundaries.  GMC has been shown to be more computationally efficient 
  than finite elements for modeling fiber composites.  A reformulated version of 
  GMC~\cite{Pindera99} has been used for the calculations in this research.

  \subsubsection{Recursive Cell Method (RCM)}
\begin{figure}[htb]
     \centering
     \scalebox{0.35}{\includegraphics{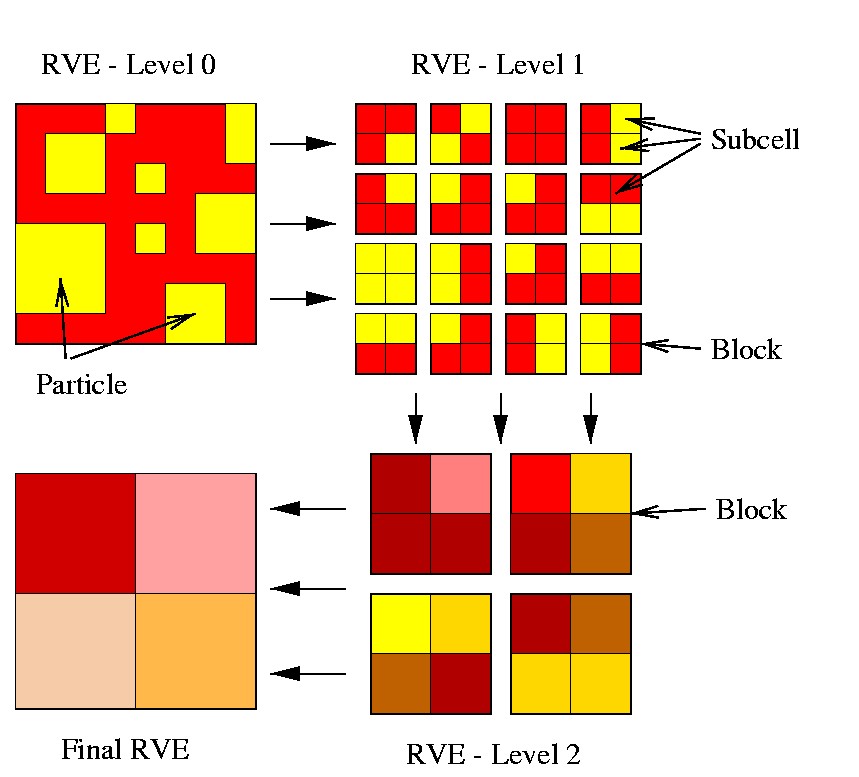}}
     \caption{Schematic of the Recursive Cell Method.}
     \label{fig1}
  \end{figure}

  The GMC technique underestimates the effective properties of PBX 9501.  The Recursive Cell
  Method (RCM) has been developed to provide a computationally efficient and accurate
  alternative to GMC.  A schematic of RCM is shown in Figure~\ref{fig1}.  The RVE is discretized 
  into subcells as in GMC.  However, instead of calculating effective properties of the whole RVE 
  in a single step, the effective properties of small blocks of subcells are determined at a time.  
  The effective properties of the RVE are calculated by combining the effective properties of blocks 
  using a recursive process.  The effective properties of each block of subcells may be determined 
  using any accurate numerical technique.  We use a finite element based technique in this 
  research.  The RCM recursive scheme has been found to reduce the computational cost and remedy 
  the shear-coupling problem of GMC.  

\section{Simulation of PBX 9501 Microstructure}
\begin{figure}[htb]
     \centering
     \scalebox{0.20}{\includegraphics{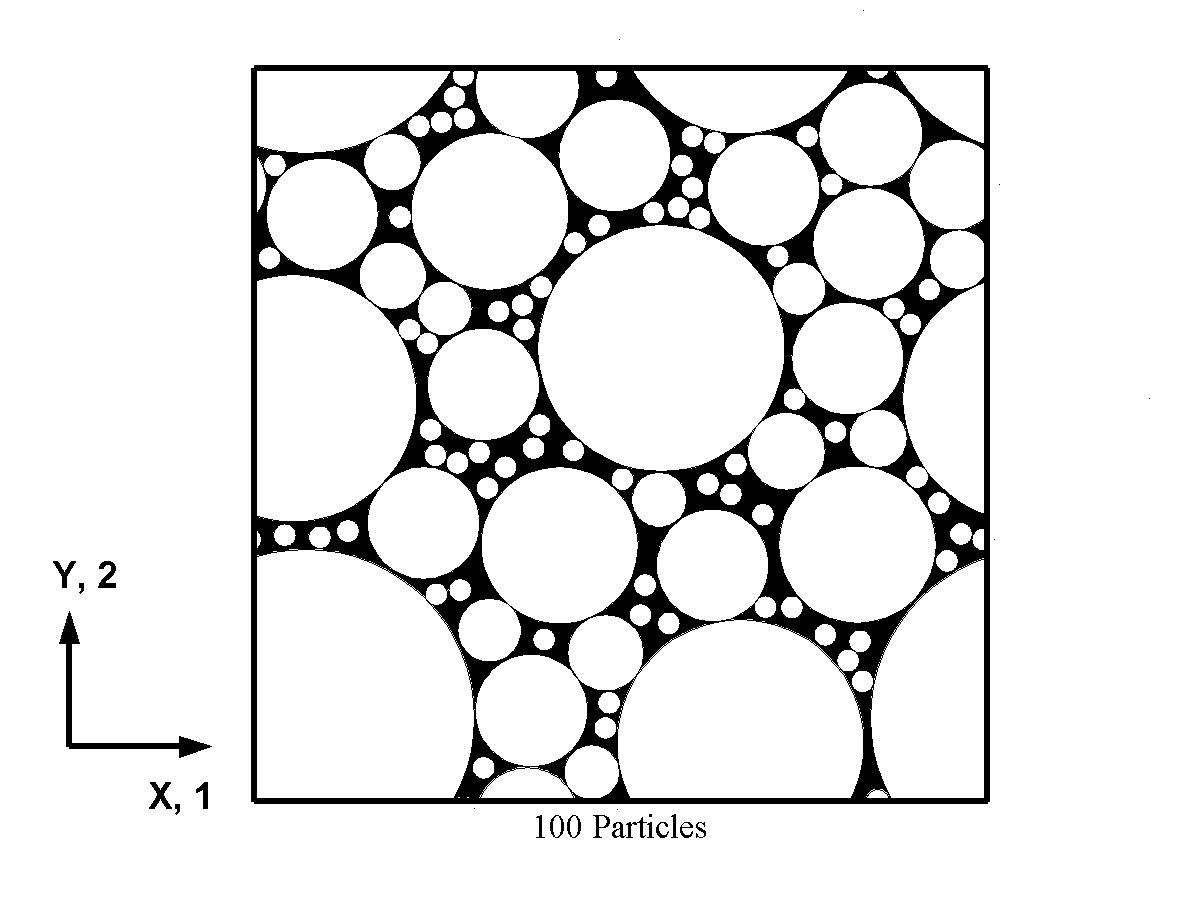}}
     \caption{Sample simulated PBX 9501 microstructure.}
     \label{fig2}
  \end{figure}

  Random sequential packing techniques were used to generate five different microstructures 
  based on the particle size distribution in the dry blend of PBX 9501~\cite{Skid97}.
  Each simulated microstructure contains around 100 particles that occupy approximately 87\% of 
  the volume of the square RVE, as shown in Figure~\ref{fig2}.  Since particles occupy 92\%
  of the volume in PBX 9501, the binder is assumed to be ``dirty'', i.e., it contains 40\% 
  HMX particles by volume so that the total volume fraction of HMX in the composite is 92\%.  The 
  effective properties of the dirty binder were calculated using the differential effective 
  medium (DEM) approach.

  The two-dimensional plane strain effective stress-strain relation for the composite is
  given by
  \begin{equation}
     \begin{bmatrix} \sigma_{11} \\ \sigma_{22} \\ \tau_{12} \end{bmatrix} = 
     \begin{bmatrix} C^*_{11}&C^*_{12}&0 \\ C^*_{12}&C^*_{22}&0 \\ 0&0&C^*_{66} \end{bmatrix}
     \begin{bmatrix} \epsilon_{11} \\ \epsilon_{22} \\ \gamma_{12} \end{bmatrix} 
  \end{equation}
  The effective stiffness matrix terms ($C^*_{ij}$) of the five PBX 9501 microstructures
  have been calculated using finite elements (FEM), GMC, and RCM.  The average effective properties
  for the five microstructures and the corresponding standard deviations are shown in 
  Table~\ref{tab5}.  On average, the five microstructures exhibit square symmetry.  The FEM 
  calculations (with $256\times 256$ square elements) 
  overestimate the effective normal stiffness ($C^*_{11}$ and $C^*_{22}$) of PBX 9501 
  by around 15\% to 25\% on average.  The corresponding effective stiffness matrix terms 
  obtained by RCM (with $8\times8$ blocks, each containing $32\times32$ subcells) are 
  around 60\% higher than the FEM values  and around two times the experimental values 
  for PBX 9501.  On the other hand, the values obtained from GMC form the normal stiffness
  terms are less than 1/10th the FEM and the experimental values.  The FEM calculations
  underestimate the values of $C^*_{12}$ for PBX 9501 by around 40\%.  RCM calculates the 
  same average value of this component of the stiffness matrix as FEM.  However, GMC predicts
  a much lower value that is around 1/7th of the stiffness of PBX 9501.  The shear 
  stiffness component ($C^*_{66}$) of PBX 9501 is calculated quite accurately by FEM and
  the error is only around 10\%.  RCM underestimates this stiffness component by around
  60\%.  However, the value predicted by GMC is around 1/300 the experimental value and
  quite close to the Hashin-Shtrikman lower bound on the shear modulus.
  \begin{table}[htb]
     \caption{Effective stiffness of simulated PBX 9501 microstructures.}
     \label{tab5}
     \centering
     \small
     \renewcommand{\arraystretch}{1.25}
     \begin{tabular}{l|cc|cc|cc|cc}
	\hline\hline
             & $C^*_{11}$ & Std. Dev. & $C^*_{22}$ & Std. Dev.& $C^*_{12}$ & Std. Dev.& $C^*_{66}$ 
		& Std. Dev. \\
	     & (GPa)      & (GPa)    & (GPa)      & (GPa)    & (GPa)      & (GPa)    & (GPa) 
                & (GPa) \\
	\hline
        PBX 9501& 1.60 &      & 1.60 &      & 0.88 &      & 0.38 &  \\
	\hline
        FEM  & 2.03 & 0.81 & 1.85 & 0.51 & 0.54 & 0.20 & 0.51 & 0.34 \\
        RCM  & 3.20 & 0.96 & 2.99 & 0.50 & 0.54 & 0.27 & 0.27 & 0.16 \\
        GMC  & 0.15 & 0.02 & 0.15 & 0.02 & 0.12 & 0.01 & 0.01 & 0.001 \\
	\hline\hline
     \end{tabular}
  \end{table}

\section{Discussion}
  The FEM calculations on the simulated PBX 9501 microstructures show that
  the microstructures are nearly isotropic.  A better approximation of isotropy can be
  obtained if larger RVEs are modeled.  The variability between models can also be
  reduced with larger RVEs.  However, larger RVEs require larger meshes to model the
  increased degrees of freedom and hence are limited by computational cost.  The FEM 
  calculations are also found to predict values of stiffness that are slightly higher 
  than the experimentally determined stiffness of PBX 9501.  Three-dimensional 
  calculations could lead to the prediction of lower stiffness for the PBX 9501 
  microstructures.  In addition, the actual material contains voids and cracks that 
  reduce the stiffness.  The finite element calculations could incorporate 
  interface elements to model interfacial debonds.  

  The RCM calculations can be performed at least three times faster than the full 
  finite element calculations.  Higher computational speeds can be obtained by reducing 
  the number of subcells per block at each level of recursion though there is some loss
  in accuracy.  Though the normal stiffness matrix components predicted by RCM are
  higher than those predicted by FEM, the values are quite acceptable considering the
  large modulus contrast between the components of PBX 9501.  Additionally, the
  RCM calculations provide much better estimates of the effective properties than GMC,
  analytical models or rigorous bounds at relatively low computational cost.

  The low effective normal stiffness predicted by GMC is because stress bridging effects are 
  underestimated by this technique.  The low shear stiffness is due to the lack of coupling between 
  the normal and shear stresses and strains in GMC.  Modifications of GMC that attempt
  to rectify these problems have been found to lose the computational efficiency
  of the original formulation.

\section{Conclusion}
  The effective properties of the models of PBX 9501 are quite close to experimentally 
  determined properties of PBX 9501.  Therefore model RVEs using circular particles
  distributed according the particle distribution of a dry blend of PBX 9501 can be
  use to simulate PBX 9501.  The RCM technique can be used to calculate effective
  properties of particulate composites quite accurately and at low computational cost.  
  The technique is much faster than full FEM calculations using the same amount of 
  discretization.  Improved performance and accuracy can be obtained in RCM by proper 
  choice of the number of subcells per block at each level of recursion.  GMC is 
  inadequate for calculating the effective properties of high energy materials because
  of the underestimation of stress bridging effects and the lack of coupling between 
  normal and shear stresses and strains.
  
\section{Acknowledgments}
  This research was supported by the University of Utah Center for the Simulation of
  Accidental Fires and Explosions (C-SAFE), funded by the Department of Energy,
  Lawrence Livermore National Laboratory, under subcontract B341493.

%
%
%
\appendix
%
%
\bibliography{em2002paper}

\begin{thebibliography}{}

\bibitem[\protect\citeauthoryear{Aboudi}{Aboudi}{1996}]{Aboudi96}
Aboudi, J. (1996).
\newblock ``Micromechanical analysis of composites by the method of cells -
  update.''\ {\em Appl. Mech. Rev}, 49(10), S83--S91.

\bibitem[\protect\citeauthoryear{Berryman and Berge}{Berryman and
  Berge}{1996}]{Berry96}
Berryman, J.~G. and Berge, P. (1996).
\newblock ``Critique of two explicit schemes for estimating elastic properties
  of multiphase composites.''\ {\em Mech. Mater.}, 22, 149--164.

\bibitem[\protect\citeauthoryear{Buryachenko}{Buryachenko}{2001}]{Burya01}
Buryachenko, V.~A. (2001).
\newblock ``Multiparticle effective field and related methods in micromechanics
  of composite materials.''\ {\em Appl. Mech. Rev.}, 54(1), 1--47.

\bibitem[\protect\citeauthoryear{Day, Snyder, Garboczi, and Thorpe}{Day
  et~al.}{1992}]{Day92}
Day, A.~R., Snyder, K.~A., Garboczi, E.~J., and Thorpe, M.~F. (1992).
\newblock ``The elastic moduli of a sheet containing circular holes.''\ {\em J.
  Mech. Phys. Solids}, 40(5), 1031--1051.

\bibitem[\protect\citeauthoryear{Gibbs and Popolato}{Gibbs and
  Popolato}{1980}]{Gibbs80}
Gibbs, T.~R. and Popolato, A. (1980).
\newblock {\em LASL Explosive Property Data}.
\newblock Univ. of California Press, Berkeley, California.

\bibitem[\protect\citeauthoryear{Gray~III, Idar, Blumenthal, Cady, and
  Peterson}{Gray~III et~al.}{1998}]{Gray98}
Gray~III, G.~T., Idar, D.~J., Blumenthal, W.~R., Cady, C.~M., and Peterson,
  P.~D. (1998).
\newblock ``High- and low-strain rate compression properties of several
  energetic material composites as a function of strain rate and
  temperature.''\ {\em Proc., 11th International Detonation Symposium},
  Snowmass, Colorado.  76--84.

\bibitem[\protect\citeauthoryear{Greengard and Helsing}{Greengard and
  Helsing}{1998}]{Greeng98}
Greengard, L. and Helsing, J. (1998).
\newblock ``On the numerical evaluation of elastostatic fields in locally
  isotropic two-phase composites.''\ {\em J. Mech. Phys. Solids}, 46,
  1441--1462.

\bibitem[\protect\citeauthoryear{Hashin}{Hashin}{1962}]{Hashin62}
Hashin, Z. (1962).
\newblock ``The elastic moduli of heterogeneous materials.''\ {\em J. Appl.
  Mech.}, 29, 143--150.

\bibitem[\protect\citeauthoryear{Hashin}{Hashin}{1983}]{Hashin83}
Hashin, Z. (1983).
\newblock ``Analysis of composite materials - a survey.''\ {\em J. Appl.
  Mech.}, 50, 481--505.

\bibitem[\protect\citeauthoryear{Hashin and Shtrikman}{Hashin and
  Shtrikman}{1963}]{Hashin63}
Hashin, Z. and Shtrikman, S. (1963).
\newblock ``A variational approach to the theory of the elastic behavior of
  multiphase materials.''\ {\em J. Mech. Phys. Solids}, 11, 127--140.

\bibitem[\protect\citeauthoryear{Markov}{Markov}{2000}]{Markov00}
Markov, K.~Z. (2000).
\newblock ``Elementary micromechanics of heterogeneous media.''\ {\em
  Heterogeneous Media : Micromechanics Modeling Methods and Simulations}, K.~Z.
  Markov and L. Preziosi, eds., Birkhauser, Boston,  1--162.

\bibitem[\protect\citeauthoryear{Michel, Moulinec, , and Suquet}{Michel
  et~al.}{2001}]{Michel01}
Michel, J.~C., Moulinec, H., , and Suquet, P. (2001).
\newblock ``A computational scheme for linear and non-linear composites with
  arbitrary phase contrast.''\ {\em Int. J. Numer. Meth. Engng.}, 52, 139--160.

\bibitem[\protect\citeauthoryear{Milton}{Milton}{1981}]{Milton81}
Milton, G.~W. (1981).
\newblock ``Bounds on the electromagnetic, elastic and other properties of
  two-component composites.''\ {\em Phys. Rev. Lett.}, 46(8), 542--545.

\bibitem[\protect\citeauthoryear{Milton}{Milton}{2002}]{Milton02}
Milton, G.~W. (2002).
\newblock {\em Theory of Composites}.
\newblock Cambridge University Press.

\bibitem[\protect\citeauthoryear{Mishnaevsky and Schmauder}{Mishnaevsky and
  Schmauder}{2001}]{Mish01}
Mishnaevsky, L.~L. and Schmauder, S. (2001).
\newblock ``Continuum mesomechanical finite element modeling in materials
  development: A state-of-the-art review.''\ {\em Appl. Mech. Rev.}, 54(1),
  49--66.

\bibitem[\protect\citeauthoryear{Pindera and Bednarcyk}{Pindera and
  Bednarcyk}{1999}]{Pindera99}
Pindera, M.-J. and Bednarcyk, B.~A. (1999).
\newblock ``An efficient implementation of the generalized method of cells for
  unidirectional, multi-phased composites with complex microstructures.''\ {\em
  Composites: Part B}, 30, 87--105.

\bibitem[\protect\citeauthoryear{Rosen and Hashin}{Rosen and
  Hashin}{1970}]{Rosen70}
Rosen, B.~W. and Hashin, Z. (1970).
\newblock ``Effective thermal expansion coefficients and specific heats of
  composite materials.''\ {\em Int. J. Engng. Sci.}, 8, 157--173.

\bibitem[\protect\citeauthoryear{Skidmore, Phillips, Son, and Asay}{Skidmore
  et~al.}{1997}]{Skid97}
Skidmore, C.~B., Phillips, D.~S., Son, S.~F., and Asay, B.~W. (1997).
\newblock ``Characterization of hmx particles in pbx 9501.''\ {\em Proc.,
  Topical Conference on Shock Compression of Condensed Matter}, American
  Institute of Physics, Los Alamos, New Mexico.  112--119.

\bibitem[\protect\citeauthoryear{Torquato}{Torquato}{2001}]{Torquato02}
Torquato, S. (2001).
\newblock {\em Random Heterogeneous Materials : Microstructure and Macroscopic
  Properties}.
\newblock Springer-Verlag, New York.

\bibitem[\protect\citeauthoryear{Zaug}{Zaug}{1998}]{Zaug98}
Zaug, J.~M. (1998).
\newblock ``Elastic constants of $\beta$-hmx and tantalum, equations of state
  of supercritical fluids and fluid mixtures and thermal transport
  determinations.''\ {\em Proc., 11th International Detonation Symposium},
  Snowmass, Colorado.  498--509.

\end{thebibliography}
%
%
%
\end{document}